\documentclass{elsart41}
\usepackage{graphics}
\usepackage{graphicx}
\usepackage{amssymb}
\begin{document}

\begin{frontmatter}


\title{Spin-charge-orbital ordering on triangle-based lattices}
\author{Hiroaki Onishi\corauthref{cor1}},
\author{Takashi Hotta}

\address{%
Advanced Science Research Center,
Japan Atomic Energy Research Institute,
Tokai, Ibaraki 319-1195, Japan}

\corauth[cor1]{%
Corresponding author.\\
\textit{Email address:} onishi@season.tokai.jaeri.go.jp (H. Onishi)}

\begin{abstract}

We investigate the ground-state property of
an $e_{\rm g}$-orbital Hubbard model at quarter filling on a zigzag chain
by exploiting the density matrix renormalization group method.
When two orbitals are degenerate,
the zigzag chain is decoupled to a doble-chain spin system
to suppress the spin frustration
due to the spatial anisotropy of the occupied orbital.
On the other hand, when the level splitting is increased
and the orbital anisotropy disappears,
a characteristic change in the spin incommnsurability is observed
due to the revival of the spin frustration.

\end{abstract}

\begin{keyword}
Geometrical frustration
\sep Orbital ordering
\sep Spin incommensurability
\sep Density matrix renormalization group method
\PACS 75.10.-b; 71.10.Fd; 75.30.Et
\end{keyword}

\end{frontmatter}


The magnetic property of frustrated spin systems has been
one of the central issues for many years
in the research field of condensed matter physics
\cite{ref-Diep-review}.
For example, it is well known that
in the antiferromagnetic (AFM) Ising model
on a triangular lattice,
there occurs macroscopic degeneracy for possible spin configurations
in the ground state
\cite{ref-Wannier-TI}.
In general, however, such high degeneracy is lifted
to suppress the spin frustration,
since the lattice is deformed
to lower the lattice symmetry
due to the spin-lattice coupling.

On the other hand,
recently there has been a rapid increase of interest in the effect of
the interplay of spin and orbital degrees of freedom
\cite{ref-Dagotto-review}.
It has been emphasized that the orbital anisotropy
plays a significant role to cause a variety of cooperative phenomena
in realistic materials.
In particular,
in geometrically frustrated lattices,
it is expected that
orbital ordering occurs to affect
the spin frustration due to the spin-orbital coupling,
since the orbital anisotropy leads to the non-uniform exchange interactions.

In this paper,
to clarify the key role of the orbital anisotropy
in geometrically frustrated lattices,
we investigate an $e_{\rm g}$-orbital Hubbard model
on a zigzag chain with one electron per site (quarter filling).
When the Hund's rule coupling $J$ is small,
the ground state is found to be paramagnetic (PM)
\cite{ref-Onishi-1,ref-Onishi-2},
which is relevant to a geometrically frustrated antiferromagnet.
Here we study the property of the PM phase
and set $J$=0 for simplicity.
The effect of $J$ has been investigated
for an $e_{\rm g}$-orbital degenerate model
\cite{ref-Onishi-2}.

The Hamiltonian considered here is given by
\begin{eqnarray}
 H &=&
 \sum_{{\bf i},{\bf a},\gamma,\gamma',\sigma}
 t_{\gamma\gamma'}^{\bf a}
 d_{{\bf i}\gamma\sigma}^{\dag} d_{{\bf i}+{\bf a}\gamma'\sigma}
 -(\Delta/2) \sum_{\bf i}
 (\rho_{{\bf i}a}-\rho_{{\bf i}b})
 \nonumber\\
 &&
 +U \sum_{{\bf i},\gamma}
 \rho_{{\bf i}\gamma\uparrow} \rho_{{\bf i}\gamma\downarrow}
 +U' \sum_{{\bf i}} 
 \rho_{{\bf i}a} \rho_{{\bf i}b},
\end{eqnarray}
where $d_{{\bf i}a\sigma}$($d_{{\bf i}b\sigma}$)
is the annihilation operator for an electron with spin $\sigma$
in the 3$z^2$$-$$r^2$($x^2$$-$$y^2$) orbital at site ${\bf i}$,
$\rho_{{\bf i}\gamma\sigma}$=%
$d_{{\bf i}\gamma\sigma}^{\dag}d_{{\bf i}\gamma\sigma}$,
and
$\rho_{{\bf i}\gamma}$=%
$\sum_{\sigma}\rho_{{\bf i}\gamma\sigma}$.
$t_{\gamma,\gamma'}^{\bf a}$ is nearest-neighbor hopping
between $\gamma$ and $\gamma'$ orbitals
along the ${\bf a}$ direction.
Note that the zigzag chain is considered as a double chain
connected by a zigzag path.
The hopping amplitudes are given by
$t_{aa}^{\bf x}$=$t/4$,
$t_{ab}^{\bf x}$=$t_{ba}^{\bf x}$=$-\sqrt{3}t/4$,
$t_{bb}^{\bf x}$=$3t/4$
for the double-chain direction and
$t_{aa}^{\bf u}$=$t/4$,
$t_{ab}^{\bf u}$=$t_{ba}^{\bf u}$=$\sqrt{3}t/8$,
$t_{bb}^{\bf u}$=$3t/16$
along the zigzag path.
Hereafter, $t$ is taken as the energy unit.
$\Delta$ is the level splitting
between 3$z^2$$-$$r^2$ and $x^2$$-$$y^2$ orbitals,
$U$ is the intraorbital Coulomb interaction,
and $U'$ is the interorbital Coulomb interaction.

We analyze the model with $N$ sites in the open boundary condition
by using the density matrix renormalization group (DMRG) method
\cite{ref-Schollwock-review}.
To reduce the size of the superblock Hilbert space,
we treat each orbital as a site.
We employ the finite-system algorithm
with keeping up to 200 states per block
and the truncation error is estimated to be
4$\times10^{-6}$ at most.

\begin{figure}[t]
\begin{center}
\includegraphics[width=0.45\textwidth]{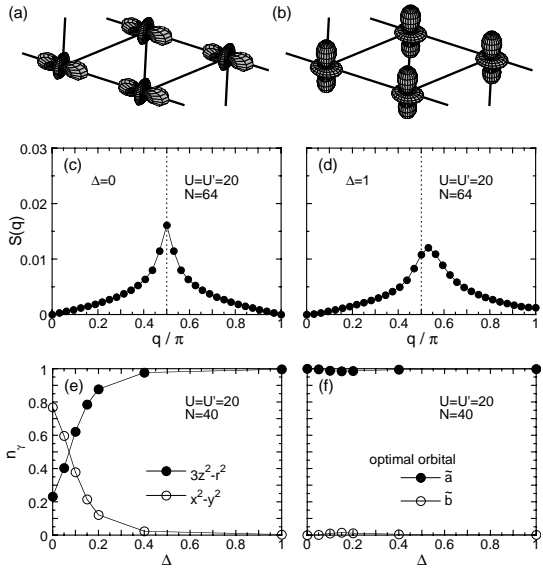}
\end{center}
\caption{%
Orbital structure for (a) $\Delta$=0 and (b) $\Delta$=1
from the DMRG results for $N$=40 at $U'$=$U$=20.
Spin correlation for (c) $\Delta$=0 and (d) $\Delta$=1.
Electron densities in
(e) 3$z^2$$-$$r^2$ and $x^2$$-$$r^2$ orbitals
and (f) optimal $\tilde{a}$ and $\tilde{b}$ orbitals.}
\end{figure}

In order to determine the orbital structure,
we introduce new operators by using an angle $\theta_{\bf i}$ such as
$\tilde{d}_{{\bf i}\tilde{a}\sigma}=
e^{i\theta_{\bf i}/2}
[\cos(\theta_{\bf i}/2)d_{{\bf i}a\sigma}+
\sin(\theta_{\bf i}/2)d_{{\bf i}b\sigma}]$
and
$\tilde{d}_{{\bf i}\tilde{b}\sigma}=
e^{i\theta_{\bf i}/2}
[-\sin(\theta_{\bf i}/2)d_{{\bf i}a\sigma}+
\cos(\theta_{\bf i}/2)d_{{\bf i}b\sigma}]$
\cite{ref-Hotta-berryphase}.
The optimal orbitals, $\tilde{a}$ and $\tilde{b}$, are determined
so as to maximize the orbital correlation
$T({\bf q})=
\sum_{{\bf i},{\bf j}}
e^{i{\bf q}\cdot({\bf i}-{\bf j})}
\langle \tilde{T}_{\bf i}^{z}\tilde{T}_{\bf j}^{z} \rangle/N^2$
with
$\tilde{T}_{\bf i}^z=
\sum_{\sigma}
(\tilde{d}_{{\bf i}\tilde{a}\sigma}^{\dag}
\tilde{d}_{{\bf i}\tilde{a}\sigma}-
\tilde{d}_{{\bf i}\tilde{b}\sigma}^{\dag}
\tilde{d}_{{\bf i}\tilde{b}\sigma})/2$.
In Figs.~1(a) and (b),
the orbital structure is shown for $\Delta$=0 and 1, respectively.
When two orbitals are degenerate for $\Delta$=0,
orbital degree of freedom is active,
but a 3$x^2$$-$$r^2$ orbital is selectively occupied
to suppress the spin frustration, as shown in Fig.~1(a).
Namely, the orbital shape extends
just along the double-chain direction,
and the zigzag chain is decoupled to a double-chain spin system
due to the orbital anisotropy.
In fact, the ratio of the AFM exchange interaction
along the double-chain direction $J_2$ to that along the zigzag path $J_1$
is estimated as $J_2/J_1$=$64^2$.
On the other hand,
for $\Delta$=1,
a lower-energy 3$z^2$$-$$r^2$ orbital is favorably occupied,
as shown in Fig.~1(b).
Note that when the 3$z^2$$-$$r^2$ orbital is fully occupied
for infinite $\Delta$,
the orbital anisotropy disappears in the $xy$ plane,
i.e., $J_2/J_1$=1,
and the spin frustration becomes effective.

In accordance with the variation in the orbital shape,
the spin state is also changed.
To clarify this point, it is convenient to reduce the present model
to a spin system on the orbital-ordered background.
Then, the present system is described by the zigzag spin chain,
in which the spin correlation has a commensurate peak
at $q$=$\pi$ for 0$\leq$$J_2/J_1$$\leq$1/2, but the peak
is gradually changed to an incommensurate one for
$J_2/J_1$$\geq$1/2, and eventually,
we find the incommensurate peak at $q$=$\pi/2$
for infinite $J_2/J_1$
\cite{ref-Tonegawa-zigzag,ref-White-zigzag}.
In Fig.~1(c), we show our DMRG result of the spin correlation
$S({\bf q})=
\sum_{{\bf i},{\bf j}}
e^{i{\bf q}\cdot({\bf i}-{\bf j})}
\langle S_{\bf i}^{z}S_{\bf j}^{z} \rangle /N^2$
with
$S_{\bf i}^z=
\sum_{\gamma}
(\rho_{{\bf i}\gamma\uparrow}-\rho_{{\bf i}\gamma\downarrow})/2$
for $\Delta$=0.
We find a clear peak at $q$=$\pi/2$,
consistent with that of the zigzag spin chain with large $J_2/J_1$,
since $J_2/J_1$=$64^2$ for $\Delta$=0.
On the other hand, as shown in Fig.~1(d),
the peak position for $\Delta$=1 changes
from $q$=$\pi/2$ toward $q$=$\pi$,
expected by analogy with the zigzag spin chain,
since we estimate $J_2/J_1$=1.61 for $\Delta$=1.
The detail of the $\Delta$ dependence
will be discussed elsewhere in future.

Finally, let us consider how the orbital state changes
in the intermediate region.
In Fig.~1(e), we show the $\Delta$ dependence of the electron densities
$n_{\gamma}=\sum_{\bf i}\langle\rho_{{\bf i}\gamma}\rangle/N$
in 3$z^2$$-$$r^2$ and $x^2$$-$$r^2$ orbitals.
With increasing $\Delta$, electrons are forced to accommodate
in the lower 3$z^2$$-$$r^2$ level, but the electron density
in each orbital is found to change gradually
without any singularity.
To understand this behavior,
we evaluate the electron densities
for the optimal orbitals.
As shown in Fig. 1(f),
it is found that one of the optimal orbitals is occupied
irrespective of $\Delta$
and the fluctuation is very small even in the intermediate region.
Namely, the present system is always regarded as a one-orbital system,
although we have considered the multi-orbital system.

In summary, for $\Delta$=0, the 3$x^2$$-$$r^2$ orbital
is selectively occupied to suppress the spin frustration
and the zigzag chain is decoupled to a double chain
due to the orbital anisotropy.
For large $\Delta$, the 3$z^2$$-$$r^2$ orbital is occupied
and the spin frustration revives, leading to the change
in the spin commensurability.

T.H. is supported by the Japan Society for the Promotion of Science
and by the Ministry of Education, Culture, Sports, Science,
and Technology of Japan.



\end{document}